# THz Dynamics of Nanoconfined Water by Ultrafast Optical Spectroscopy

A. Taschin[1], P. Bartolini[1] and R. Torre [1,2],

[1] European Laboratory for Non-Linear Spectroscopy (LENS),
Via N. Carrara 1, Sesto Fiorentino, 50019 Firenze, Italy.
Email: torre@lens.unifi.it

[2] Dip. di Fisica ed Astronomia, Università di Firenze
Via G. Sansone 1, Sesto Fiorentino, 50019 Firenze, Italy.

*ABSTRACT*

*We investigated the vibrational dynamics and the structural relaxation of water nanoconfined in porous silica samples with pore size of 4 nm at different levels of hydration and temperature. We used as spectroscopic technique the time-resolved optical Kerr effect, which enables to investigate the ultrafast water dynamics in a wide time (0.1-10 picosecond) or frequency (10-0.1 THz) window. At low levels of hydration, corresponding to two complete superficial water layers, no freezing occurs and the water remains mobile at all the investigated temperatures, while at the fully hydration we witness to a partial ice formation at about 248 K that coexists with the part of surface water remaining in the supercooled state. At low hydration, both structural and vibrational dynamics show significant modifications compared to the bulk liquid water due to the strong interaction of the water molecules with silica surfaces. Inner water, instead, reveals relaxation dynamics very similar to the bulk one.*

## 1 INTRODUCTION

The confinement of a liquid in a solid matrix having micro/nanometric cavities is relevant to understand a wide range of systems, both natural and artificial. The porous rocks and the cement/concrete are two immediate examples. In a nanoconfined liquid a significant fraction of molecules are in direct contact with the confining surfaces; these interactions can produce a non-trivial modification of the structural and dynamic properties of the liquid. The liquid water is likely the most relevant substance that is frequently subject to nanoconfinement conditions; often, silica or silicate materials constitute the solid matrix. The interaction of water molecules with silica surfaces is particularly strong and stable and it is responsible for the substantial modification of the hydrogen bond networks of the liquid. These studies are also useful to enlighten the complex physics of bulk water, which remains a much-debated issue [1].

We investigated a prototype system, made by a porous silica matrix with controlled pore dimensions of 4 nanometers filled by liquid water [2, 3, 4, 5]: Vycor 7030. This porous glass is a well-studied and characterized matrix, which enables reliable experimental studies. We studied the structural and dynamic features of nanoconfined water by non-linear time-resolved spectroscopy, optical Kerr effect that allowed the measurement of the water dynamics in a wide temporal range, from tens of femtoseconds (few THz in the frequency window) to tens of picoseconds (tenths of THz). The used experimental set-up employs a specific polarization control of the laser pulses that provides a double improvement; it enables a very stable optical heterodyne detection and an automatic suppression of the non-phase sensitive components of the signal. With these improvements, together with a precise measurement of the instrumental response, we achieved an excellent signal/noise statistics and we obtained a reliable extraction of the material response function both in time and frequency domains [6].

We investigated the properties of nanoconfined water at variable hydration level and temperature; we proved that a substantial fraction of water, the "outer" water corresponding to about 50%, does show a partial distortion of the hydrogen bond networks and a slowed dynamics. Moreover, this water fraction does not crystallize even at low temperature remaining in a liquid-like phase. The remaining "inner" part of water show structural and vibrational dynamics similar to bulk water.

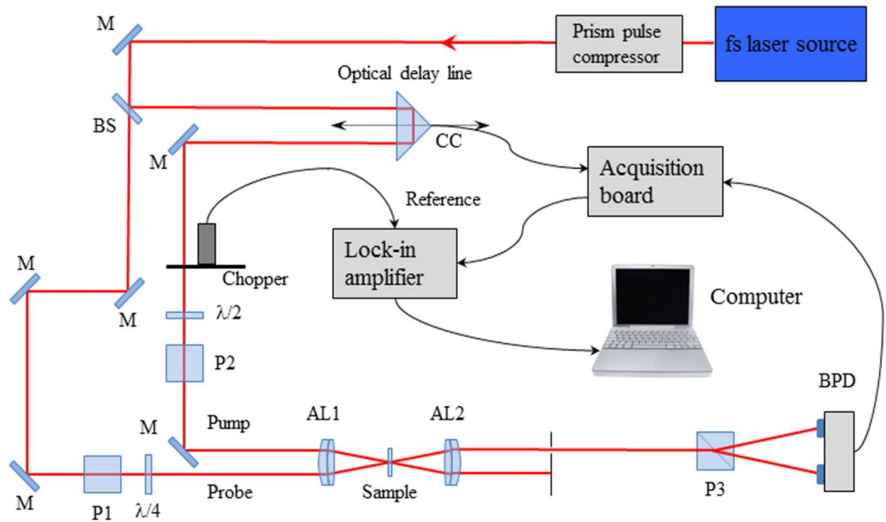

**Fig. 1:** Optical set-up for optical Kerr effect spectroscopy: M mirror, BS beam splitter, CC corner cube, λ/2 and λ/4 half and quarter wave plate, P# polarizers, AL# achromatic lenses, and BPD the balance photodiode.

## 2 EXPERIMENTAL FEATURES

### 2.1 Optical Kerr effect experiment

The Heterodyne Detected Optical Kerr Effect (HD-OKE) experiment probes the relaxation of a birefringence induced by a short pump laser pulse in the sample through third-order non-resonant effects. The birefringence relaxation processes are defined by the matter response function that contains the structural and vibrational dynamics [6, 7]. The time-resolved OKE technique is known for many years to be a valid spectroscopic tool to measure the complex dynamics in condensed matter [8, 9, 10]. Recently, the new femtosecond laser sources made the HD-OKE a reliable and successfully spectroscopy, that enables a detailed analysis of the measured response functions [11, 12, 13].

Exhaustive description of the theoretical features and experimental details about the experiment are reported in the references [6, 7]. Fig. 1 shows a sketch of the HD-OKE experiment set-up. The laser pulses are produced by a Ti:Sapphire Kerr-lens mode locking cavity and their group velocity dispersion is controlled by a prism compression stage. A beam splitter (BS) divides the laser beam in the pump and probe beams. The probe pulse is delayed respect to the pump pulse by an optical delay line whose translation stage is computer controlled. The half wave plate (λ/2) and P2 polarizer fix the pump linear polarization at 45° respect to vertical direction (i.e. orthogonal to the experimental plane); the P1 polarizer and quarter wave plate (λ/4) sets the probe to have a circular polarization. Probe and pump beams are focused inside the sample by the achromatic lens AL1. The probe and signal beams are then collected by the achromatic lens

AL2 and send to the Wollaston P3 polarizer. This selects and splits the two polarization components present in the beam, horizontal or vertical. Each polarization component is given by the sum of the OKE signal with the residual probe (i.e. local field), which has opposite phase in the vertical respect to the horizontal component; so the heterodyne contribution has opposite sign in the two measured components. The balanced photodiode detector BPD makes an electronic subtraction of the two polarization components extracting the HD-OKE signal from the other common-mode contributions. The output of the photodiode goes to a lock-in amplifier together with the reference signal of the pump chopper. A DAQ acquisition board simultaneously acquires the translation stage position from the encoder board and the lock-in output signal. Finally, the computer stores the signal-positions and reproduces the signal decay.

The OKE signals in water or confined water are typically very low and fast, thus precise determination of the instrumental function is extremely important in order to extract the sample response. In our experiments the instrumental function was obtained by measuring the HD-OKE signal in a reference sample characterized by a simple and well-known nuclear response. As reference sample we chosed a plate of calcium fluoride (CaF2) which has only one Raman active band in the probed frequency range (322 cm$^{-1}$) [6].

## 2.1 Samples preparation

We used Vycor 7930 to nanoconfine water; This is an open cell porous glass, which is produced by a spinodal demixing in a borate-rich and borate-poor glass and subsequent bleaching of the borate-rich phase. It has nominal values of porosity and mean pore size of 28% and 4 nm respectively. The solid constituent of Vycor 7930 is the glass Vycor 7913, which is composed of 96% of silica, of 3% of boron oxide and the remaining part mostly of aluminum oxide and zirconium oxide. Vycor is strongly hydrophilic because of the numerous silanol groups present on the widespread inner surfaces (around 250 m$^2$/g).

Our sample has been purchased from ''Advanced Glass and Ceramics'' in the shape of square slab with side of 10 mm and thickness of 2 mm (Fig. 2). These were cleaned by immersion in 35% hydrogen peroxide solution at 90 °C for 2h and, after repeated washing in distilled water, the samples were stored in P$_2$O$_5$ (phosphoric anhydride) until usage. Dry Vycor samples were obtained by heating the slabs at 400 °C for 10 h, while samples at different hydration level have been prepared by exposing dry Vycor slabs at water atmosphere for 12 h in closed vials. Samples, with different amount of water filling the nano-pores, were obtained via a vapor phase exothermic transfer from the bulk. The water concentration in each sample was adjusted adding different volume of bulk water. We prepared samples at six different hydration levels, whose water content was accurately determined from their weights: $f = 0.1\%$, $f = 3.3\%$, $f = 5.6\%$, $f = 8.5\%$, $f = 10.5\%$ and $f = 24.3\%$ where $f$ is the filling fraction parameter, which is given by the ratio between the weight of the contained water and dry Vycor weight. We can consider the sample at the lowest hydration as substantially "dry", in fact at this hydration the number of water molecules inside the pores is much less than the number of silanol groups present on the pore surface (5 OH per nm$^2$ [4,14]). At the hydration $f = 5.6\%$, there is enough water inside pores to have the full coverage of the pore surface with a monomolecular water layer, "1-layer". The sample with 10.5% hydration contains about "2-layers", whereas the sample with 24.3% hydration is in the "full hydration" condition. Each sample, after the hydration procedure, was placed in a cylindrical gold-plated copper cell and it was cooled by a closed cycle helium cryostat with a temperature stability of ± 0.1 K.

We used Fourier Transform Infrared (FT-IR) spectroscopy to characterize the hydration process of porous glass. Spectra were recorded in the range of 4200-5500 cm$^{-1}$, where two absorption bands can be well-distinguished at ∼ 4550 and ∼ 5260 cm$^{-1}$ arising from combination of stretching and bending modes of silanol and water, respectively. Fig. 3 shows the measured FT-IR spectra as function of hydration. The comparison between the absorption spectra of water (∼ 5260 cm$^{-1}$) is reported in the left panel and clearly indicates that the structure of water drastically changes as a function of water content. The band maximum of water, see right panel of Fig.3, is red-shifted with increasing water concentration, while a broad absorption grows below 5200 cm$^{-1}$, which gradually approaches the spectral profile of bulk water. The sharp peak at ∼5260 cm$^{-1}$ is reasonably assigned to vibrations of monomeric water, whilst the growing broad red-shifted band to the H-bonding interactions, moreover this latter band is present even at the lowest water concentration (<5%) suggesting that water aggregation occurs before full surface coverage is attained.

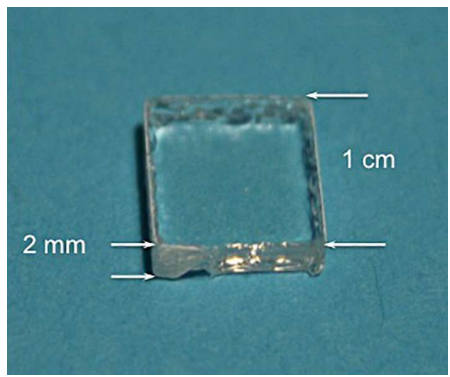

**Fig. 2:** Vycor 7930 sample used for confining water.

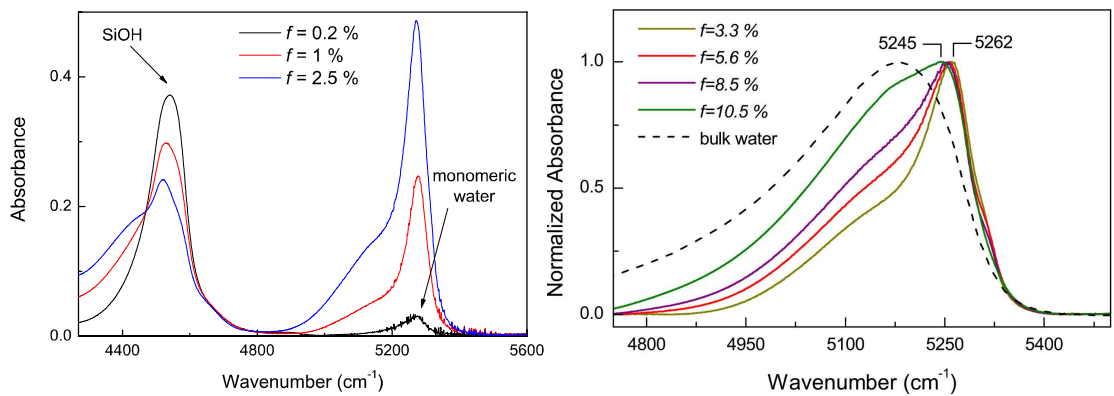

**Fig. 3:** Fourier Transform Infrared spectra of the Vycor-water sample as function of the hydration level. Left panel: extended frequency range where both silanol and physical water bands are measured. Right panel: absorbance spectra of the water band on partially filled Vycor matrices at different hydration levels.

## 2 RESULTS

HD-OKE data have been measured for all the six hydrations quoted in the previous section at the room temperature of 293 K. Here we focused on three hydration levels $f$ = 5.6 %, 11.6% and 24.3%, corresponding to 1-layer, 2-layers and full hydrations, respectively. For the 2-layers and full hydration levels, the data were also collected at different temperatures from 293 K down to 248 K. Below this temperature a partial crystallization occurs in the full hydration sample.

## 2.1 Data analysis in the time-domain

The data analysis performed in the time-domain enables a direct and valuable examination of the slower dynamics, corresponding to the structural relaxation processes. In order to reproduce the signals in the time-domain and extract the relaxation dynamics parameters, we used the following expressions [4, 5]:

$$S(t) = \int [k\delta(t-t') + R_n(t-t')] G(t') dt' \quad (1)$$

$$R_n(t) = AR_{dry}(t) + B\frac{d}{dt}\exp\left[-\left(\frac{t}{\tau}\right)^\beta\right] + \sum_i C_i \exp(-\gamma_i^2 t^2)\sin(\omega_i t) \quad (2)$$

where $R_n(t)$ is the third-order nuclear response function and $G(t)$ the instrumental function [6]. In the Eq. (2), the $R_{dry}(t)$ describes the response of the dry matrix and it has been simulated by an appropriate function found by the fit of the dry sample signal; The second term accounts for the structural relaxation processes; The last term for the fast oscillating/vibrational contributions present in the signal.

The slow dynamics was analyzed fitting the data with the previous expressions. The stretched exponential (summed to the dry contribution) reproduces very well the long part of the data, but due to the complexity of the data we were not able to extract reliable values for the stretching parameter, $\beta$, which has been fixed to the bulk water value of 0.6 [1].

In Fig. 4 we reports the measured data at changing of hydration with their fits for the temperature of 293 K. In the figure, we report also the HD-OKE data for bulk water at the same temperature. It is evident that water confined response reveals several differences compared to the bulk water one. At short times, the fast oscillations typical of the water cage dynamics are smoothed and it is present a very fast vibrational contribution due to the Vycor silica signal. At long time, a slow oscillating contribution is added to the structural relation response. This slow oscillation is present also in the signal of dry Vycor ($f$ = 0.1%) and is probably due to an acoustic-like vibration localized on the pore surface [15].

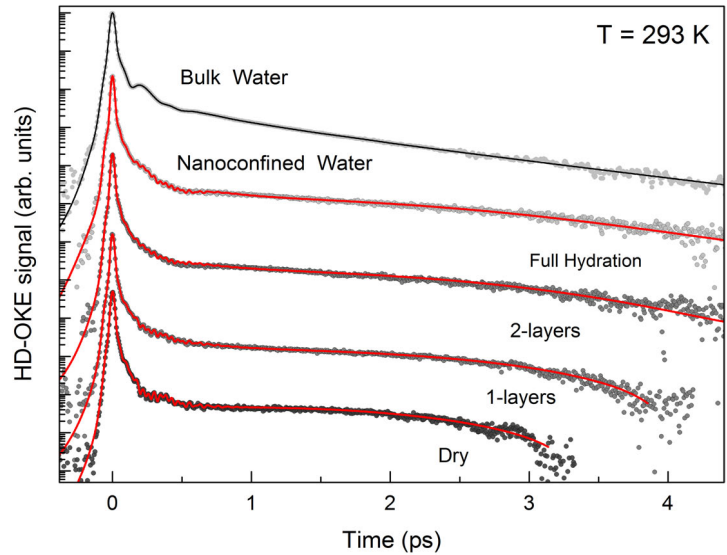

**Fig. 4:** HD-OKE data of bulk water and nanoconfined water at different hydration levels for T=293 K. Full circles: experimental data; continuous line: best-fit results. The data are reported in a log-linear plot.

In Fig. 5 we report the HD-OKE data for nanoconfined water for the 2-layers and full hydration levels at different temperatures together with the response of the dry matrix, contribution whose change with temperature is totally negligible. At a simple look these data reveals that there are no signatures of ice formation at any temperature for the sample with 2-layers, whereas the presence of ice is clear in the full hydrated sample at 248 K, the lower temperature measured. In fact, this appears as a clear under-dumped oscillation (i.e. a sharp peak at about 215 cm$^{-1}$ in the frequency domain, see afterwards) easily detectable in the data. The supercooled nanoconfined water shows dynamics apparently similar to the bulk water.

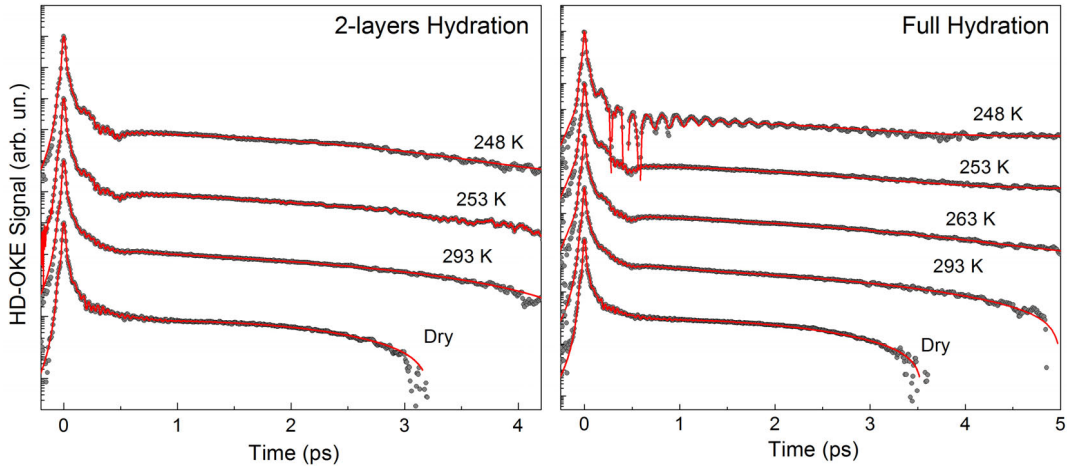

**Fig. 5:** HD-OKE data of nanoconfined water at different temperatures for the 2-layers and full hydration levels in a log-linear plot. Full circles: experimental data; continuous line: best fit results.

Fig. 6 shows the structural relaxation times as a function of the temperature for 2-layers and full hydration. For comparison the constant times found for bulk water are reported too. The structural relaxation times of nanoconfined water are comparable with the bulk one, even if some differences are present. At high temperatures, the structural relaxation times of full hydration water is indeed intermediate between the bulk and 2-layers water relaxation times; this likely due to the superposition of the contributions of the inner water (faster) and of the surface layers (slower). The 2-layers water is about a factor of 3 slower than the bulk water and shows a reduced temperature dependence. So that the relaxation time approach the bulk values at the lower temperatures. The full hydration shows an unexpected a slowing down process approaching the lower temperatures, stronger than that takes place in the bulk water. This could be due to the formations of ice crystals inside the pores at those temperatures.

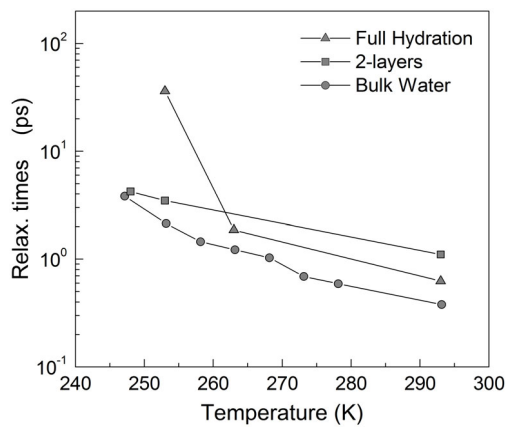

**Fig. 6:** Structural relaxation times as a function of temperature at different hydrations.

Our results suggest that the collective rearrangements of interfacial and inner water are characterized by similar structural processes (both require a stretched exponential function to be properly reproduced), but with clearly different characteristic time scales. The dynamics of the inner water cannot be measured selectively; nevertheless, from our results we can argue that its relaxation times must be very close to those of bulk water.

## 2.1 Data analysis in the frequency-domain

The vibrational dynamics can be better discussed in the frequency domain. The spectra can be obtained performing the Fourier Transform (*FT*) of the measured data, deconvoluted from the instrumental response, and retaining the imaginary part of it [6]:

$$Im[\tilde{R}_n(v)] \propto Im\{FT[S(t)]/FT[G(t)]\} \qquad (3)$$

All the spectra show a common band at 24 THz [4, 5]. This band is related to the optical phonon typical of the amorphous $SiO_2$ [16] and it is present in all the measured signals with same intensity independently of the hydration and the temperature values. We used this band to perform a reliable normalization. In order to extract the spectra of the nanoconfined water, the spectrum of the dry Vycor matrix, $f = 0.1$ %, were subtracted from the complete experimental spectra.

In Fig. 7 we show the spectra of the nanoconfined water at the fixed temperature of 293 K for samples with different hydrations. Fig. 8 reports the frequency spectra for the two different hydrations: 2-layers (right panel) and full (left panel) at decreasing of temperature. Always for comparison, frequency responses of bulk water and hexagonal ice are shown in the right and the left panel, respectively.

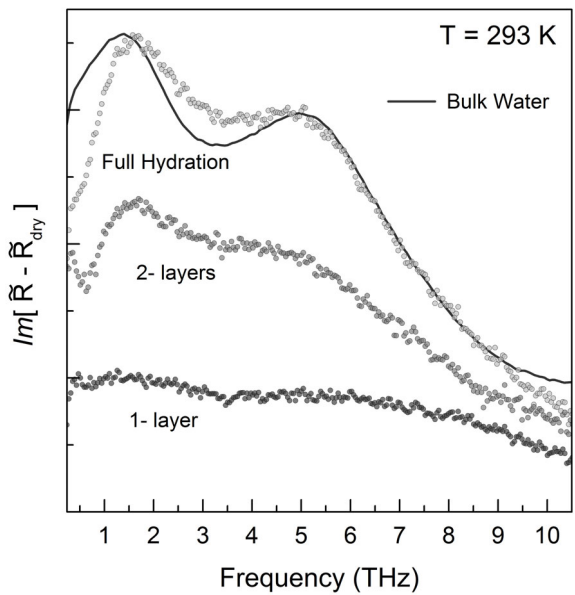

**Fig. 7:** Fast Fourier Transform (FFT) of HD-OKE data, subtracted from the dry Vycor contribution, at changing of the hydration and fixed temperature. For comparison, the FFT of bulk water at the same temperature is reported too.

The nanoconfined water shows fast dynamics that is strongly dependent by the hydration level. In fact, the 1-layer spectrum does not show the bending and stretching bands that are typical for bulk water. These bands are intermolecular vibrations involving the H–O⋯H hydrogen bonds. These experimental data proves that in the 1-layer water the H-bond networks is strongly distorted by the surfaces interactions. Increasing the hydration level, we observe the gradual formation of these water intermolecular bands suggesting that the H-bond network start to be similar to the bulk water system.

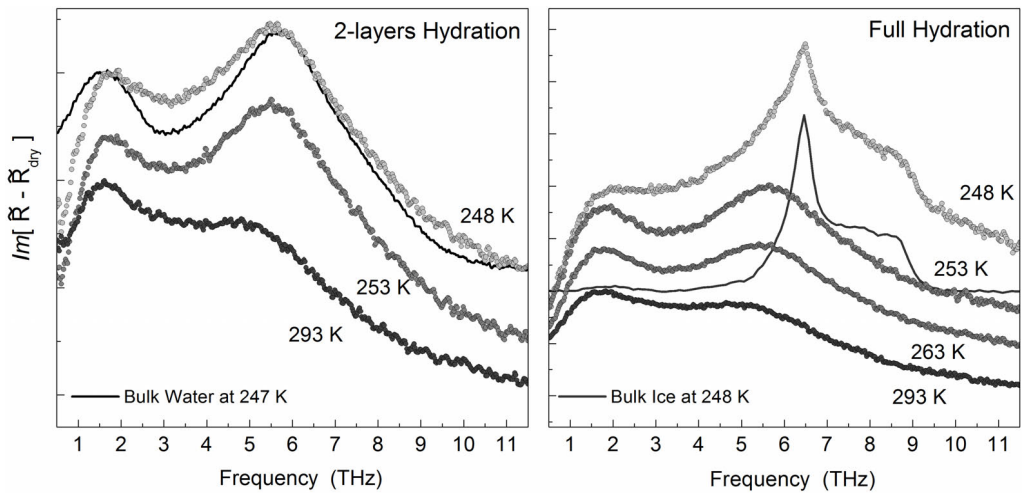

**Fig. 8:** Fast Fourier Transform the HD-OKE data, subtracted from the dry Vycor contribution, at different temperatures for the two hydration levels. For comparison, spectrum of bulk water is shown in the right panel and that of the hexagonal ice in the left panel. The data has been shifted vertically for better view; The bulk water and ice data has been shifted by the same values used for the nanoconfined data at 248 K.

The spectrum of nanoconfined water at 2-layers hydration show a significant temperature dependence; in particular, the bending and stretching bands develop rapidly and become very similar to the bulk one at 248 K. Moreover, we do not observe any ice formation in the 2-layers nanoconfined water, whereas in the full hydrated sample at 248 K the vibrational spectrum clearly shows the presence of ice into nanopores. Nevertheless, a comparison with the vibrational spectrum of bulk ice undoubtedly evidences that not all the nanoconfined water crystalize. In fact, there is an extra broad vibrational component, extending over the whole frequency window. In our opinion, this is due to those parts of water that remain mobile and it retains its vibrational dynamics and a finite structural relaxation time. A comparison of the reported experimental spectra suggests that the amount of water remaining in a supercooled state is a non-negligible part of the nanoconfined water. We estimated it from the spectrum area to be about 50% of the total amount of internal water. This is in fairly good agreement with the volumes occupied by the outer water (i.e. the two water layers at the pore surface) and the inner water (i.e. the remaining 4–5 layers of water filling the internal part of the pore).

## 3 CONCLUSION

The HD-OKE data of nano-confined water show a complex structural and dynamics scenario. Nevertheless, our results show some useful evidences.

The data at increasing hydration suggests that the water filling is proceeding as a layer-by-layer process; they can be easily rationalize according to this hypothesis even if we cannot exclude completely the presence of water plugs (i.e. water bridge across the pore surfaces) [17].

We found the 2-layers hydration level to be a threshold beyond which water starts to have bulk like characteristics. The HD-OKE data at variable temperature strengthen this interpretation. In fact, the 2-layers hydration water does not crystallize at 248 K preserving a liquid-like structure; this due to the interaction with the silica surfaces that strongly modifies the water networks up to the second layer. Lowering the temperature both the structural, see Fig. 6, and the vibrational dynamics, see Fig. 8, of the 2-layers water become closer to the bulk water, suggesting that the interactions with the surfaces become less effective respect to the water-water H-bonds.

The features of nano-confined water at full hydration resemble that of bulk water; both the structural relaxation and the vibrational spectrum are similar to the bulk. Below temperature of 248 K freezing occurs, but not all the nano-confined water crystallizes; from the measured spectrum, we estimated that about 50% of water remains liquid that corresponds to about the two outer water layers.

The present results, at changing of hydration and temperature, can be rationalized considering the presence of two water types inside the hydrophilic nano-pores: the "outer water", i.e. the water close to pore surface, and the "inner water", i.e. the part of the water inside the pore (see Fig. 9). We must distinguish the outer water, i.e. water in contact/close to silica surfaces at full hydration, from the multilayer water, i.e. water close to surfaces at partial hydration. The numerical simulation can selectively investigates the outer/inner water components [18,19], whereas the experimental investigations can measure only the multilayer water.

Considering a van der Waals diameter for a water molecule of the order of 0.3 nm, the thickness of the first two water layers covers about 1.2 nm of the pore diameter. Since, the Vycor pore diameter is 4 nm there are about 4–5 layers of inner water. The geometrical estimation of the volumes are 50% for each water forms. Our HD-OKE results on water confined in hydrophilic nano-pores support the existence of the outer and inner waters. Moreover, our data suggest a high similarity between the outer and 2-layers water.

We would point out how a pore diameter smaller than 4 nm involves a higher amount of outer water then inner. For diameter smaller 1.5 nm practically the nano-confined water is totally of the outer type, and the water interactions with the pore surfaces determine the structural and dynamic processes.

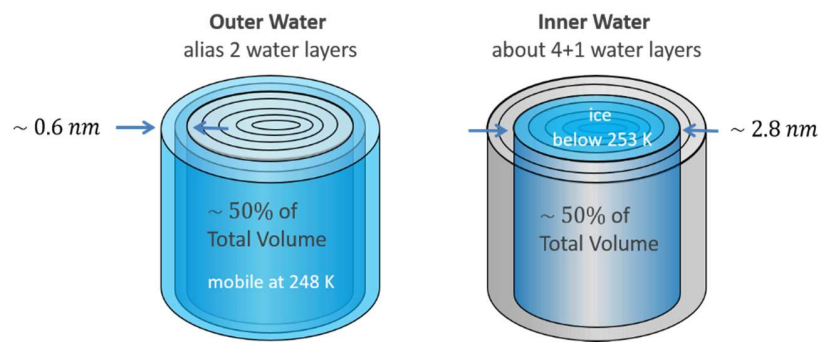

**Fig. 9:** Different types of water in the silica nanoporous